\documentclass[%
 reprint,
superscriptaddress,
 amsmath,amssymb,
 aps,
]{revtex4-2}

\usepackage{graphicx}
\usepackage{dcolumn}
\usepackage{bm}
\usepackage{float}

\begin{document}

\preprint{APS/123-QED}

\title{Non-Reciprocal Transport of Thermally-Generated Magnons}


\author{M. Cosset-Ch\'eneau}
\email{m.n.c.g.cosset-cheneau@rug.nl}
\affiliation{%
 Zernike Institute for Advanced Materials, University of Groningen, 9747 AG Groningen, The Netherlands}
 \author{S.H. Tirion}
\affiliation{%
 Zernike Institute for Advanced Materials, University of Groningen, 9747 AG Groningen, The Netherlands}
 \author{X.Y. Wei}
\affiliation{%
 Zernike Institute for Advanced Materials, University of Groningen, 9747 AG Groningen, The Netherlands}
\author{J. Ben Youssef}
\affiliation{%
Lab-STICC, CNRS, Université de Bretagne Occidentale, Brest, France}
\author{B. J. van Wees}
\affiliation{%
 Zernike Institute for Advanced Materials, University of Groningen, 9747 AG Groningen, The Netherlands}
\date{\today}

\begin{abstract}
We demonstrate the non-reciprocity of electrically and thermally-generated incoherent magnon transport using the magnetization direction of a Py wire placed on top of an ultratin YIG film. We show that the transport properties of thermally-generated magnons under a Py wire depends on the relative orientation between the temperature gradient and the Py-magnetization direction. The symmetries of this non-reciprocal magnon transport match with those predicted by the remote dipolar interaction between YIG and Py magnons, controlled by the chirality of the YIG magnon dipolar stray fields. We also show that the directional magnon generation by the spin Seebeck effect from the Py wire displays the symmetries expected from the chiral spin Seebeck effect.

\end{abstract}

\maketitle

\section{Introduction}

The transport of information using spin angular momentum is one of the important aims of spintronics, which has been achieved using the electrons spin \cite{ZuticRMP2004}. Spin waves of angular momentum, or magnons, have also been demonstrated to be efficient information carriers over long distances in magnetic insulators \cite{CornelissenNP2015,LebrunN2018}. In contrast to electronic spins, magnons can generate a dynamic dipolar stray field \cite{BertelliSA2020}, which provides an additional knob to control their generation and propagation \cite{BertelliAQT2021, MaePRB2022, QinNC2021,ChenJPDAP2021}. 
This can be achieved in thin magnetic films thanks to the remote dipolar coupling between magnons in adjacent magnetic layers \cite{YuPR2023,WangNR2020}. Such a coupling depends on the chirality of the magnon dipolar stray fields linked to their propagation direction, and on the magnetization direction of the magnetic films \cite{YuPRL2019}. 
Recent experiments studying single frequency coherent spin waves in YIG emitted and detected by microwave striplines have shown a non-reciprocal transmission and directional emission of magnons far from equilibrium, attributed to this remote magnon-magnon coupling. It was in particular observed that the transmission of magnons in YIG under an array of ferromagnetic wires depends on the wire magnetization direction \cite{ChenPRB2019,MucchiettoACSN2024}, and that the emission direction of coherent magnons at radiofrequencies in YIG by a ferromagnetic wire also depends on its magnetization direction \cite{SimonNL2019,WangPRB2023}.

The efficient use of magnons as information carriers in spintronics would benefit from their generation by a d.c. electrical current. This has been achieved using nonlocal magnon transport devices consisting of two heavy metal strips, an injector and a detector, deposited on a ferromagnetic insulator \cite{CornelissenNP2015}. The application of a bias current in the injector results in a nonlocal signal in the detector owing to near-equilibrium diffusive and incoherent magnon transport. %
The spin Hall effect in the injector creates a magnon accumulation or depletion in the ferromagnetic insulator with respect to the magnon population at equilibrium \cite{BenderPRL2012}. This accumulation (or depletion) is described by a magnon chemical potential \cite{cornelissenPRB2016} which varies spatially in the ferromagnetic insulator depending on its magnon transport and relaxation properties \cite{WeiNM2022}. It is measured by the detector through the inverse spin Hall effect. The magnon chemical potential generated by spin-charge interconversion in the injector will be referred to as being electrically-generated. %

The bias current in the injector also results in a temperature gradient in the ferromagnetic insulator by Joule heating. 
This temperature gradient generates a flow of thermally-generated magnons, creating a magnon accumulation under the detector \cite{ShanPRB2016,ShanPRB2017}, detected by spin-charge interconversion processes, and referred to as the nonlocal spin Seebeck effect \cite{KikkawaARCMP2023}. This thermally-generated signal therefore depends on thermal transport parameters and on the magnon transport and relaxation properties \cite{bauerNM2012,anPRB2021}.

The observation of a non-reciprocal transport and directional generation of coherent magnons indicates that similar effects could be achieved using the dipolar magnon-magnon coupling for electrically and thermally-generated incoherent magnons in a YIG film, on which a ferromagnetic strip has been placed.
However, in contrast to the case of coherent magnons, the signal measured in nonlocal magnon transport devices originates from a deviation of the magnon population with respect to its equilibrium value.
In linear response, the non-reciprocity of the nonlocal signal can therefore not be obtained from an average magnon flow direction. Only the position of the magnon injector and detector for electrically-generated magnon, or of the heating point and detector for thermally-generated magnons, are relevant for the dependence of the nonlocal signal on the magnetization configuration of the device.
In the following we call a transport process non-reciprocal if the nonlocal signal does not have the same amplitude when interchanging the magnon injection or heating point and the magnon detection point,
without changing the magnetization configuration of the device. We stress that following this definition, a non-reciprocal transport process can still verify the Landauer-Buttiker \cite{ButtikerIBM1988} or Onsager-Casimir \cite{CasimirRMP1945} reciprocity relations.

 Such a non-reciprocity of the electrically-generated magnon transport in YIG has been achieved by inserting a Py strip in between the injecting and detecting heavy metal strips \cite{HanNL2021}. The non-reciprocal thermally-generated magnon transport originating form the dipolar magnon-magnon coupling has however not been demonstrated, despite the possibility to realize thermal magnon transistor using a ferromagnetic gate \cite{YuArxiv2023}, and the predicted directional thermal generation of magnons from a ferromagnetic strip \cite{YuPRL2019}. 

 In this paper we use nonlocal magnon transport devices [Fig. 1(a)] to demonstrate the non-reciprocity of the thermally-generated magnon transport in YIG controlled by the magnetization orientation of a Py wire placed between two Pt strips. The effect of the Py magnetization direction on the nonlocal spin Seebeck effect depends on the temperature gradient direction, reversed when interchanging the heating and magnon detection points, and on the YIG magnetization direction. This indicates that it originates from the interaction between magnons in Py and in YIG, which is controlled by the chirality of the magnon dipolar stray fields. We finally show that an directional nonlocal spin Seebeck effect can be generated from a Py wire, owing to this remote magnon-magnon interaction.

\section{Experimental details}
\begin{figure}[h!]
  \centering
  \includegraphics[width=8.5cm]{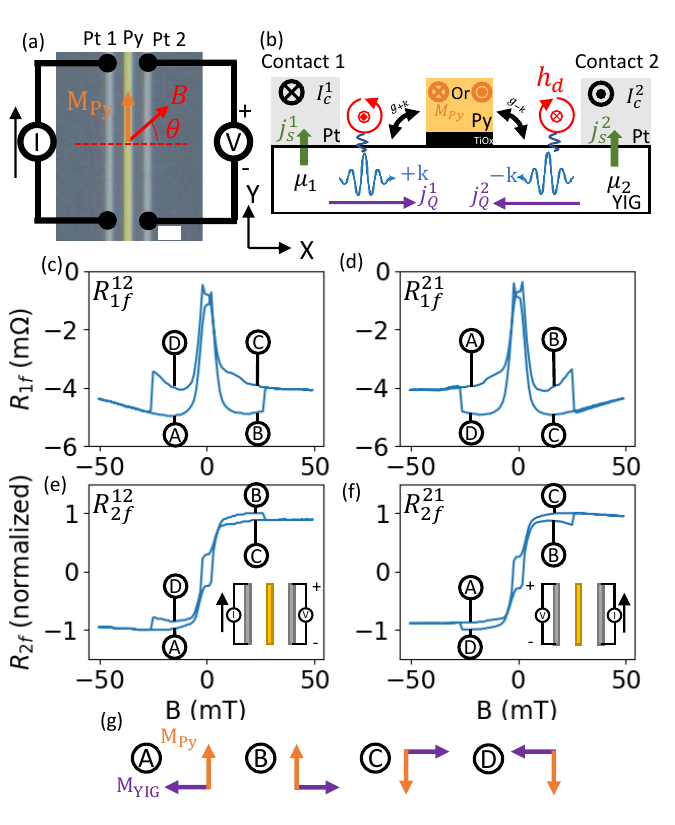}
  \caption{(a) Device geometry with the Pt wires in gray and the Py gate in yellow.
  The displayed measurement configuration corresponds to the one used for the measurement of $R^{12}_{1f}$ and $R^{12}_{2f}$.
  The white scale bar corresponds to a distance of 2 $\mu$m. (b) Lateral view of the device geometry, with the rotating dipolar stray field $\mathbf{h}_d$ generated on top of YIG by the right and left moving magnons. A bias current $I_c^i$ applied in the platinum contact $i$ ($\mathrm{i}=1,2$) creates a spin current $J_s^i$ which generates an accumulation $\mu_i$ of electrically-generated magnons in YIG, which diffuse toward the platinum contact j. In addition, the temperature gradient induced by the bias current forms a flow $J_Q^i$ of thermally-generated magnons in YIG toward the contact $j$. The electrically and thermally generated magnons both induce a spin accumulation $\mu_j$ under the contact $j$. (c) and (e) First and second harmonic signal corresponding to $R^{12}_{1f}$ and $R^{12}_{2f}$ respectively, measured in the configuration given in the inset of (e). (d) and (f) same as (c) and (d) measured in the configuration given in the inset of (f) corresponding to $R^{21}_{1f}$ and $R^{21}_{2f}$ respectively. The second harmonic signals have been normalized by their maximun values. The bias current for the measurement in (c-f) is 1000 $\mu$A, and the magnetic field angle is $\theta=-4^\circ$. (g) Py and YIG magnetic configurations corresponding to the states A-D in (c-f).}
  \label{fig:fig1}
\end{figure}

The nonlocal magnon transport devices were fabricated by electron beam lithography on 7.9 nm thick YIG ($\mathrm{Y_2Fe_5O_{12}}$) with (111) orientation deposited by liquid phase epitaxy on a GGG ($\mathrm{Gd_3Ga_5O_{12}}$) substrate \cite{WeiNM2022}. Two 7 nm thick, 400 nm wide and 80 $\mu$m long Pt strips with a center to center separation of 3 $\mu$m were dc sputtered. A 40 nm thick, 400 nm wide and 80 $\mu$m long Py strip was dc sputtered between the Pt strips, with a TiOx insertion between YIG and Py. The TiOx insertion was obtained by depositing 3 nm of Ti, oxidized before the Py deposition. It prevents the exchange coupling between YIG and Py \cite{SM}. An optical image of the device is shown in Fig. \ref{fig:fig1}(a). The measurements of the first ($1f$) and second harmonic ($2f$) nonlocal response defined as $R_{1f}=V_{1f}/I_c$ and $R_{2f}=V_{2f}/I_c^2$ have been carried out using a conventional lock-in method with a frequency kept under 50 Hz and an applied bias current $I_c$ of 1000 $\mu$A. The electrical connections for the bias current application and voltage measurements are shown in Fig. \ref{fig:fig1}(a) and Fig. \ref{fig:fig3}(a) for magnon transport under, and emission from the Py wire, respectively. The magnon transport gives a negative first harmonic nonlocal signal for this measurement configuration \cite{CornelissenNP2015}. All the measurements were performed at room temperature.

\section{First harmonic signal}
We first applied a bias current in the Pt 1 electrode [Fig. \ref{fig:fig1}(a)] while sweeping a magnetic field kept at an angle $\theta=-4^\circ$  with respect to the $\hat{x}$-direction. The field was first swept upward from negative to positive values, and then backward. The first and second harmonic responses were simultaneously measured at the Pt 2 electrode and are shown in Fig. \ref{fig:fig1}(c) and Fig. \ref{fig:fig1}(e). The bias current application and electrical signal detection electrodes were then interchanged and the same measurement sequence was performed, with the measured responses shown in Fig. \ref{fig:fig1}(d) and Fig. \ref{fig:fig1}(f).

The measured signals display abrupt jumps at $\pm 25$ mT, that we attribute to the reversal of the Py wire $\hat{y}$-component magnetization direction. Macrospin computations assuming an in-plane magnetization were used to evaluate the $\hat{x}$ and $\hat{y}$-components of the Py wire magnetization direction \cite{SM}. The YIG magnetization aligns with the applied magnetic field above 5 mT. The low field magnetic configurations corresponding to the states A-D of Fig. \ref{fig:fig1}(c-f) are shown in Fig. \ref{fig:fig1}(g). The states (A,C) constitutes a reversed pair of magnetization configurations for which all the magnetizations and magnetic fields in the system are reversed, and so does the (B,D) pair. States pairs (A,B) and (C,D) correspond to the upward and downward magnetic field sweeps with positive and negative $\hat{y}$-component of the Py magnetization, respectively. 

We denote $R^{\mathrm{ij}}_{1f}$ and $R^{\mathrm{ij}}_{2f}$ the first and second harmonic nonlocal signals obtained by applying the bias current in Pt $i$ and detecting the nonlocal voltage in Pt $j$ (with $i,j=1,2$). The first observation is that for $\mathrm{S=A,B,C}$ and $\mathrm{D}$, $R^{12}_{1f}(\mathrm{S}) \neq R^{21}_{1f}(\mathrm{S})$, which shows a non-reciprocity of the electrically-generated magnon transport according to our definition. In addition, as visible in Fig. \ref{fig:fig1}(c) [$R_{1f}^{12}$] and \ref{fig:fig1}(d) [$R_{1f}^{21}$] the signal amplitude remains the same for the reversed magnetic configurations upon interchanging the magnon injection and detection strips: $R_{1f}^{21}(\mathrm{A/B})=R_{1f}^{12}(\mathrm{C/D})$ and $R_{1f}^{21}(\mathrm{C/D})=R_{1f}^{12}(\mathrm{A/B})$. In addition, for $\mathrm{X=A,\, B}$ we observe that $|R_{1f}^{21}(\mathrm{X})|<|R_{1f}^{12}(\mathrm{X})|$ while $|R_{1f}^{21}(\mathrm{Y})|>|R_{1f}^{12}(\mathrm{Y})|$ for $\mathrm{Y}=\mathrm{C,\, D}$. 
Finally, the relations $|R_{1f}^{12}(\mathrm{A/B})|>|R_{1f}^{12}(\mathrm{C/D})|$ and $|R_{1f}^{21}(\mathrm{A/B})|<|R_{1f}^{21}(\mathrm{C/D})|$ are observed in the first harmonic nonlocal signals.


The Figs. 1(c) and 1(d) corresponds to a four terminal electrical measurement in the linear regime with terminals $a$ and $b$ corresponding to the strip Pt 1 and terminals $c$ and $d$ to Pt 2. We therefore expect that the measured nonlocal resistance $R_{ab,cd}=V_{cd}/I_{ab}$ where the current is applied from a to b and the voltage measured from $c$ to $d$ follows the Landauer-Buttiker reciprocity relation \cite{ButtikerIBM1988}. Namely, this resistance should keep the same value when interchanging the contacts and reversing the time-reversal symmetry breaking fields $\boldsymbol{B}$, $\boldsymbol{M}_\mathrm{YIG}$ and $\boldsymbol{M}_\mathrm{Py}$: $R_{ab,cd} (\boldsymbol{B},\boldsymbol{M}_\mathrm{YIG},\boldsymbol{M}_\mathrm{Py} )=R_{cd,ab} (-\boldsymbol{B},-\boldsymbol{M}_\mathrm{YIG},-\boldsymbol{M}_\mathrm{Py})$. This is what is observed in Figs. 1(c) and 1(d) where the pairs of states (A,C) and (B,D) correspond to a reversed configurations of the time-reversal symmetry breaking fields. The Landauer-Buttiker reciprocity relations therefore hold in the first harmonic measurements.

The magnon transport in a nonlocal device can however also be viewed as a two terminal magnon transport measurement. The application of a charge current in Pt $i$ results in the formation of a magnon accumulation $\mu_i$ under this strip by the spin Hall effect, and to a magnon accumulation $\mu_j$ under Pt $j$ by magnon diffusion processes [Fig. 1(b)]. The Landauer-Buttiker formalism can therefore be applied for magnon transport by considering Pt $i$ as the magnon current source and Pt $j$ as the magnon current drain. The magnon accumulation $\mu_j$ is directly measured by the inverse spin Hall effect in Pt $j$, such that from the magnon point of view, the nonlocal electrical transport appears to be a two terminal magnon transport measurement.

\begin{figure}[h!]
  \centering
  \includegraphics[width=8.5cm]{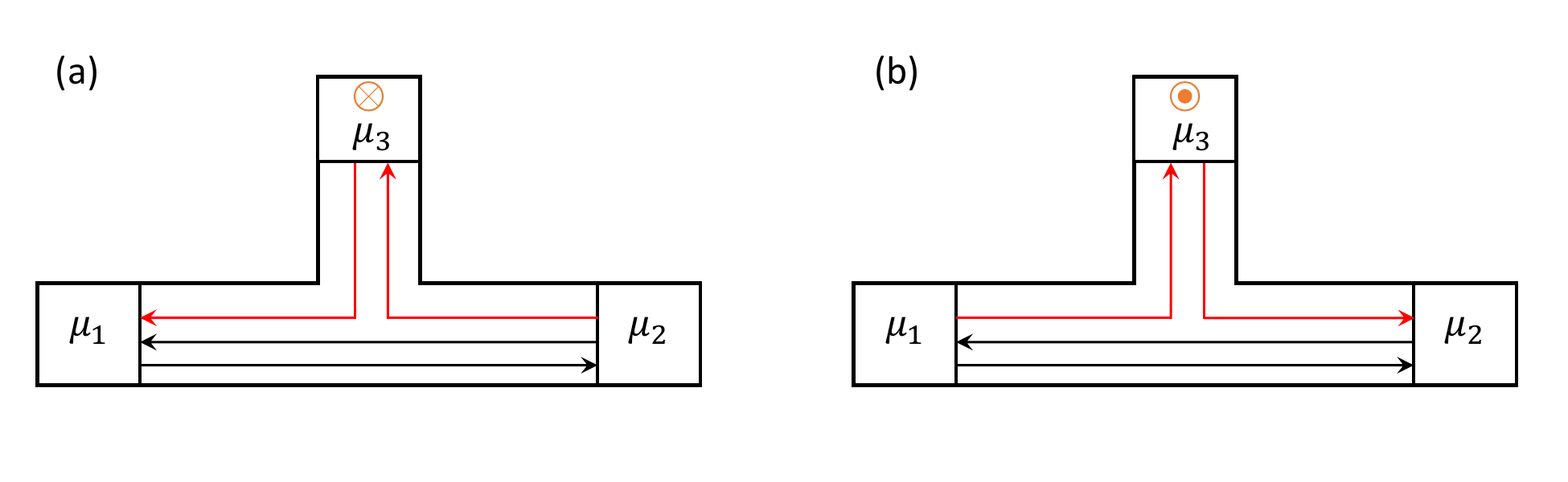}
  \caption{Magnon transmission between the Pt contacts with magnon electrochemical potentials $\mu_1$ and $\mu_2$ (black arrows), and between the Pt contacts and the Py wire with electrochemical potential $\mu_3=0$ (red arrow). The orange arrow correspond to the Py magnetization direction ($\boldsymbol{M}_\mathrm{Py}$). Individual magnons propagating from the right to the left couple differently to $\mu_3$ than those propagating from the left to the right. (a) $\boldsymbol{M}_\mathrm{Py}$ along the $+\hat{y}$ direction corresponds to the configurations A and B, while $\boldsymbol{M}_\mathrm{Py}$ along the $-\hat{y}$ direction (b) corresponds to configurations C and D.  }
  \label{fig:fig1bis}
\end{figure}

The results reported in Figs. 1(c) and 1(d) are from this point of view surprising since the Landauer-Buttiker reciprocity relations predicts that for a two terminal measurement $R_{ij} (\boldsymbol{B},\boldsymbol{M}_\mathrm{YIG},\boldsymbol{M}_\mathrm{Py} )=R_{ij} (-\boldsymbol{B},-\boldsymbol{M}_\mathrm{YIG},-\boldsymbol{M}_\mathrm{Py})$ with i the current source and j the current drain. This would correspond to, for instance, $R_{1f}^{12} (A)=R_{1f}^{12} (C)$ which is not observed in our measurements.

The only way to reconcile our first harmonic measurements with the Landauer-Buttiker reciprocity relations is to consider the measurements in Figs. 1(c) and 1(d) as a three terminal magnon transport measurements [Fig. 2]. The magnons can indeed relax in the Py wire, which in an electrical measurement would act as a third terminal connected to ground with $\mu_3=0$. The magnon transport from the source to the drain can be described as an unbalanced transmission of magnons flowing in opposite directions. These individual magnons can couple differently with the third terminal as a function of their propagation direction and the Py magnetization direction \cite{HanNL2021,YuArxiv2023}. This results in a modification of the nonlocal signal amplitude upon reversal of $\boldsymbol{M}_\mathrm{Py}$ only, or upon interchanging the magnon injection and detection points. 

In Section V we compare our results with a model proposed in Ref. \cite{YuPRL2019} which accounts to such a coupling.

\section{Second harmonic signal}

The second harmonic signals $R_{2f}^{12}$ [Fig. \ref{fig:fig1}(e)] and $R_{2f}^{21}$  [Fig. \ref{fig:fig1}(f)] have different amplitudes,
as previously observed using similar devices geometry in the absence of a middle Py wire \cite{CornelissenNP2015}. This feature is to be expected since the second harmonic response is non-linear in the applied bias current, and therefore not bound by the Onsager-Casimir reciprocity relations to display the same amplitude upon contact and magnetic fields reversal. It should however be noted that the thermodynamic force leading to the spin Seebeck effect is the temperature gradient, which should lead to a response matching the Onsager-Casimir reciprocity relations in the linear response to the thermal bias \cite{DejenePRB2014}. The asymmetries in our device, such as for instance different Pt strip resistances, however lead to different Joule heating powers and temperature gradients upon interchanging the heating and magnon detection points. Because of this, the observation of such a reciprocity is not expected in our measurements. 
 In order to study the effect of the Py wire magnetization direction on the second harmonic response upon interchanging the position of the heating point and of the detector, the signals displayed in Fig. \ref{fig:fig1}(e) and Fig. \ref{fig:fig1}(f) have therefore been normalized by their maximum values to compensate for the different temperature gradients induced by the device asymmetry.

\begin{figure}[h!]
  \centering
  \includegraphics[width=8.6cm]{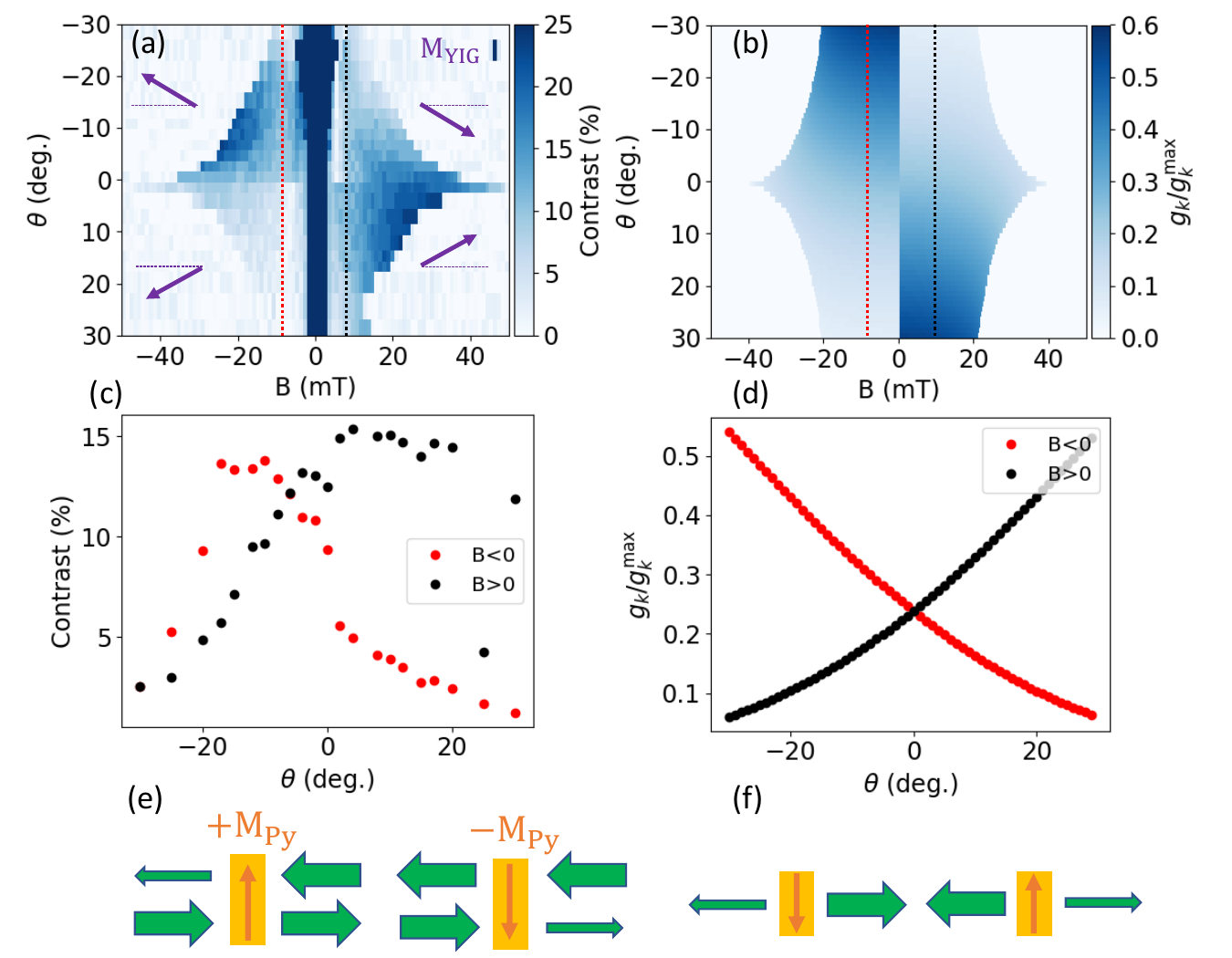}
  \caption{(a) Map of the second harmonic contrast for $R^{12}_{2f}$ as a function of the magnetic field angle and strength. The purple arrow give the YIG magnetization direction in the four quadrants. (b) Map of he computed coupling parameter normalized by its maximum value from \textbf{eq. (1)} as a function of the field angle and intensity for right moving ($+k$) magnons at 12 mT. (c) Angular dependence of the contrast for a field of 12 mT. The red and black dots correspond to the dashed lines in (a). The measurements have been performed with a 1000 $\mu$A bias current. (d) same as (c) for the computed normalized coupling parameter. (e) Schematics of the thermally-generated magnon transmission asymmetry along the $+\hat{x}$ and $-\hat{x}$ direction under the Py electrode as a function of its magnetization direction. (f) Same as (e) for the efficiency of the magnon emission in the $+\hat{x}$ and $-\hat{x}$ direction.}
  \label{fig:fig2}
\end{figure}

Several jumps are present in the second harmonic signal at field strengths corresponding the jumps observed in the first harmonic signal. Similar to the case of the first harmonic signal, $R^{12}_{2f}(\mathrm{S}) \neq R^{21}_{2f}(\mathrm{S})$ for $\mathrm{S=A,B,C,D}$ which shows that the second harmonic signal is non-reciprocal following the definition introduced above.
We observed that $|R^{12}_{2f}(\mathrm{A})|>|R^{12}_{2f}(\mathrm{D})|$ and $|R^{21}_{2f}(\mathrm{A})|<|R^{21}_{2f}(\mathrm{D})|$. The temperature gradient between the Pt strips is reversed in the two measurements configuration: $\partial_x T>0$ in the $R^{12}_{2f}$ configuration, and $\partial_x T<0$ in the $R^{21}_{2f}$ configuration. 
The thermally-generated magnon flow follows the temperature gradient, and the second harmonic nonlocal signal is proportional to the magnon accumulation or depletion caused by this flow \cite{ShanPRB2016}. The results reported in  Fig. \ref{fig:fig1}(e) and Fig. \ref{fig:fig1}(f) therefore shows that the nonlocal signal caused by the thermally-generated magnons depends on the relative orientation between the temperature gradient and the Py magnetization direction. In particular for a Py magnetization along the $+\hat{y}$ direction, the nonlocal signal generated by a temperature gradient in the $+\hat{x}$ direction is larger than for a temperature gradient in the $-\hat{x}$ direction. This relation is reversed for as Py magnetization along the $-\hat{y}$ direction. This indicates that the transmission of an individual magnon under the Py wire depends on $\boldsymbol{M}_\mathrm{Py}$ and on its propagation direction as summarized in Fig. \ref{fig:fig2}(e) [see also Fig. 2]. These results show a non-reciprocity of the transport of thermally-generated magnons, controlled by the direction of the Py magnetization.

So far we have only discussed the effect of the Py wire magnetization direction on the non-reciprocity of the nonlocal signals. In order to study the effect of the YIG magnetization angle, we performed the same measurement sequence at different magnetic field sweep angle $\theta$. 
In Fig. \ref{fig:fig2}(a) we plot for $R_{2f}^{12}$ the contrast of the nonlocal signals obtained during the upward ($\mathrm{B}_\Rightarrow$) and downward ($\mathrm{B}_\Leftarrow$) magnetic field sweep at angle $\theta$. During the upward sweep the configurations A and B are reached, while configurations C and D are reached during the downward sweep [Fig. 1 (c-f)]. The contrast is defined as $|R_{2f}^{12}(\theta,\mathrm{B}_\Rightarrow)-R_{2f}^{12}(\theta,\mathrm{B}_\Leftarrow)|/|R_{2f}^{\mathrm{max}}(\theta)| \times 100$, with $|R_{2f}^{\mathrm{max}}(\theta)|$ the maximum nonlocal signal recorded during the field sweep at angle $\theta$.  In between the Py magnetization switching events, the contrast therefore corresponds to the absolute values of the difference between the nonlocal signals measured with opposite $\hat{y}$-component of the Py magnetization. The contrast measured for the reversed contact configuration measurements is shown in the Supplemental Material \cite{SM}. 
In Fig. \ref{fig:fig2}(a), the near zero field region displays a strong contrast attributable to the switching of YIG. The field-angle map of the contrast can be divided into four regions. The positive angle-positive field and negative angle-negative field quadrants show a contrast of up to 23\%. The two other quadrants have a lower contrast of less than 10\%, with one of these quadrants having a larger contrast than the other. The angle dependence of the contrast at $\pm$ 12 mT, where the magnetization of YIG is fully saturated and the magnetization of Py is close to the $\hat{y}$-direction, is shown in Fig. \ref{fig:fig2}(c). At positive field, the contrast increases when going from negative to positive $\theta$ values, before saturating and dropping above $\theta=18^{\circ}$. The reverse $\theta$-dependence of the contrast is observed at negative field. Similar results were obtained for electrically-generated magnons \cite{SM}.


\section{Coupling between magnons in Py and YIG}

We now discuss a mechanism which could lead to the coupling discussed in Section III between the magnons in YIG and in Py. The presence of a TiOx barrier prevents an exchange coupling between YIG and Py \cite{DasPRB2020,SantosPRA2021} (see also Ref. \cite{SM}).  In addition, such non-reciprocity has not been observed in electron-based spin angular momentum transport experiments using similar magnetization configurations \cite{Cosset-CheneauPRL2021, Cosset-CheneauPRBL2022}, and in magnon transport experiments using thinner Py wires \cite{SantosPRA2021,DasPRB2020}. It is therefore likely that the observed effect originates from a magnon-specific remote interaction, such as the remote magnon-magnon coupling driven by dynamic magnon dipolar stray fields \cite{ShiotaPRL2020}, and cannot be linked with exchange interaction-driven spin or magnon relaxation mechanisms. A magnon propagating along the $\hat{x}$-direction produces a rotating dipolar stray field in the (xz) plane [Fig. \ref{fig:fig1}(b)]. These stray fields couple the magnons in YIG with the Py wire magnon bath if the handedness of their rotation matches the one of the Py magnetization \cite{kruglyakAPL2021}. The strength of the coupling is also controlled by the magnon dynamic dipolar stray field strength. This coupling results in an increase of the YIG magnon damping \cite{YuPRL2019,kruglyakAPL2021,HanNL2021}, causing a smaller signal in nonlocal magnon transport measurements. The handedness of the dipolar stray fields reverses with the magnon k-vector \cite{YuPRL2019}, hence leading to a coupling factor $g_k$ which depends on the magnon propagation direction [Fig. \ref{fig:fig1}(b)].
This would lead to a reversal of the transmission asymmetry between right and left-moving individual magnon when reversing the Py magnetization direction [Fig. \ref{fig:fig2}(e)], and provide a physical mechanism for the three terminal magnon transport measurement described in Fig. 2.
In addition, the dipolar stray field strength depends on the direction of the magnon k-vector with respect to the YIG magnetization direction \cite{YuPRL2019, HanNL2021}.  If  $\mathbf{k}\parallel M_{\mathrm{YIG}}$, there is an equal distribution of dipolar stray field strength, albeit with different rotation handedness, on both side of the YIG film. For a right moving magnon however, the dipolar stray field strength at the top YIG surface increases when $M_{\mathrm{YIG}}\cdot \hat{y} >0$ and decreases when $M_{\mathrm{YIG}}\cdot \hat{y}<0$ \cite{YuPRL2019}. This relation is reversed for a left moving magnon. 

In order to compare these predictions with our results, we evaluated the magnon-magnon coupling strength $g_k$ in our system as a function of the YIG magnetization direction. This coupling strength writes \cite{HanNL2021}:
\begin{equation}
    g_k = f(k) (1+\eta_k \mathbf{M}_{\mathrm{YIG}}\cdot\hat{y}) \times (1 - \eta_k  \mathbf{M}_{\mathrm{Py}}\cdot\hat{y})
\end{equation}
with $f(k)$ is independent of the magnetizations direction and a function of the k-vector modulus, and $\eta_{\pm k}=\pm 1 $ corresponding to a magnon flowing in the $\pm \hat{x}$ direction. This Py wire magnetization direction-dependence of the coupling matches with the measurements displayed in Fig. \ref{fig:fig1}(e) and (f), assuming that the direction of the temperature gradient and of the resulting thermally-generated magnon flow can be associated with the magnons propagation direction. 
Indeed, when the coupling is large at a given $\mathbf{M}_{\mathrm{Py}}$ an individual magnon is more likely to relax in the Py wire, thus leading to a lower transmission \cite{YuArxiv2023}. This results in a larger contrast of the nonlocal signal since $g_k$ is strongly reduced when reversing the $\hat{y}$-component of $\mathbf{M}_{\mathrm{Py}}$.
The magnetization vectors $\mathbf{M}_{\mathrm{Py}}$ and $\mathbf{M}_{\mathrm{YIG}}$ were obtained from macrospin computations \cite{SM}. The computed value of $g_{+k}$ is shown in Fig. \ref{fig:fig2}(b) and \ref{fig:fig2}(d) for comparison with the contrast measured for $R_{2f}^{12}$.  Fig. \ref{fig:fig2}(a) and \ref{fig:fig2}(b) display the same high and low intensity quadrants. In addition, the contrast measured for magnons flowing in the $-\hat{x}$ direction evolves consistently with the computation for $g_{-k}$, with an inversion of the high and low contrast quadrants \cite{SM}. The field strength and angle dependence of the contrast is therefore qualitatively consistent with what is expected from a remote magnon-magnon coupling controlled by the chirality of the magnon dipolar stray field.
The positive and negative low field contrast angle dependence in Fig. \ref{fig:fig2}(c) presents the crossing expected from the computation of $g_{+k}$ [Fig. \ref{fig:fig2}(d)]. The saturation and drop at higher magnetic field angle in Fig. \ref{fig:fig2}(c) is however not reproduced in our computations and may originate from a breakdown of the macrospin model at larger angles, with the presence of magnetic domains in the Py wire. Finally, the difference of contrast in the low contrast quadrants could be due to the magnon mode ellipticity, expected to affect the $180^{\circ}$ periodicity of the contrast \cite{YuPR2023, ChenPRB2019}. The possible role of the Py-wire static dipolar stray field on the non-reciprocity of the signal is discussed in the Supplemental Material \cite{SM}, and ruled out as they would produce signals with symmetries different to those reported in Fig. \ref{fig:fig1}(c-f).

Overall, our results are consistent with a transmission asymmetry of right and left moving magnons under the Py wire, controlled by the chirality of the magnon dipolar stray field and Py magnetization direction.
Following the interpretation provided in Ref. \cite{YuArxiv2023}, this coupling results in the flow of thermally-generated magnons to partially relax into the Py wire, hence causing the nonlocal spin Seebeck effect to depend on the temperature gradient direction and on the Py magnetization direction.

As predicted in Ref. \cite{YuPRL2019}, this remote magnon-magnon interaction is also expected to allow for the directional emission of thermally-generated magnons from the Py wire. In the following part of this paper we report experimental results consistent with this prediction.

\section{Directional Magnon Emission}
\begin{figure}[h!]
  \centering
  \includegraphics[width=8.6cm]{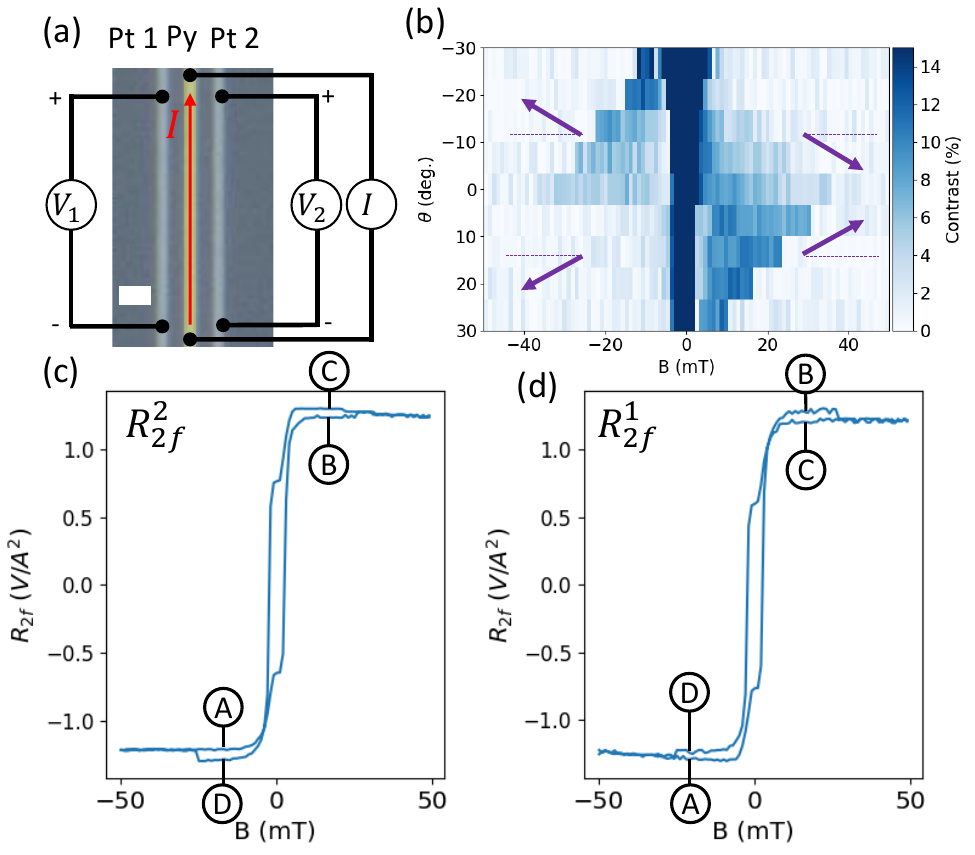}
  \caption{(a) Contact configuration for the measurement of the magnon emission from the Py gate. The white scale corresponds to a distance of 2 $\mu$m. (b) Contrast of the signal measured at Pt 2 as a function of the intensity and angle of the applied magnetic field. (c) Second harmonic nonlocal signal measured at Pt 2 ($R_{2f}^2$). (d) Same as (c), measured at Pt 1 ($R_{2f}^1$). The states A-D correspond to those reported in Fig. \ref{fig:fig1}(g). A bias current of $1000$ $\mu$A has been used for the measurements in (b) and (d), and the magnetic field angle is $\theta=-4^\circ$.}
  \label{fig:fig3}
\end{figure}

The emission of thermally-generated magnons from the Py wire was studied by flowing a charge current into the Py wire using the electrical contact configuration presented in Fig. \ref{fig:fig3}(a). 
The magnons generated by a temperature gradient in the $-\hat{x}$ and $+\hat{x}$ directions were detected through second harmonic measurements at the Pt 1 and Pt 2 electrodes respectively, and the nonlocal signal measured at Pt i is denoted $R_{2f}^{\mathrm{i}}$ ($i=1,2$). The rest of the measurement sequence is the same as in the magnon transport experiment described in Section III and IV. No first harmonic signal was observed, which is as expected since the presence of a $\mathrm{TiO_x}$ barrier between YIG and Py prevents the electrical injection of magnons in YIG. The second harmonic signals measured in Pt 2 and Pt 1 at a magnetic field angle $\theta=-4^\circ$ are shown in Fig. \ref{fig:fig3}(c) and Fig. \ref{fig:fig3}(d) respectively. We observed a signal jump pattern related to the switching of the Py wire magnetization as in Fig. \ref{fig:fig1}(c-f), with $|R_{2f}^{1}(\mathrm{D})|<|R_{2f}^{1}(\mathrm{A})|$ and  $|R_{2f}^{2}(\mathrm{D})|>|R_{2f}^{2}(\mathrm{A})|$, superimposed on the regular nonlocal spin Seebeck effect.  
The Py wire therefore emits a larger magnon flow in the $-\hat{x}$ direction  when the $\hat{y}$-component of its magnetization is positive than when it is negative. The reverse relation with the Py magnetization direction holds for the magnon flow in the $+\hat{x}$ direction. This indicates that the charge current in the Py wire emits a directional flow of thermally generated magnons, as summarized in Fig. \ref{fig:fig2}(f). 

Interestingly, the dependence of the second harmonic signal amplitude with the sign of the Py wire magnetization $\hat{y}$-component in Fig \ref{fig:fig3}(c) and \ref{fig:fig3}(d) is reversed with respect to the transmission measurements performed in Pt 1 and Pt 2, and presented in Fig. \ref{fig:fig1}(e) and \ref{fig:fig1}(f). In particular, $|R_{2f}^{2}(\mathrm{D})|>|R_{2f}^{2}(\mathrm{A})|$ [Fig. 4(c)] while $|R_{2f}^{12}(\mathrm{D})|<|R_{2f}^{12}(\mathrm{A})|$ [Fig. 1(e)].
This shows that for an individual magnon propagating in a given direction, a larger relaxation under the Py wire is correlated with a larger emission from this same Py wire [see Fig. \ref{fig:fig2}(e) and \ref{fig:fig2}(f)] .

This correlation  is consistent with predictions of microwave-generated coherent magnon directional transmission and emission from a ferromagnetic wire at radiofrequencies \cite{AuAPL2012_1,AuAPL2012_2}. It is verified in the entire accessible $\theta$ range \cite{SM}. We then measured the contrast at Pt 2 at different applied magnetic field angle $\theta$ and obtained a field-angle map of the contrast [Fig. \ref{fig:fig3}(b)] similar to the one reported in Fig. \ref{fig:fig2}(a) [see Ref. \cite{SM} for the contrast at Pt 1]. This indicate that the observed directional emission of magnons has the same physical origin as their non-reciprocal transport, that is the dipolar chiral stray field-mediated remote magnon-magnon interaction.

This effect, termed the chiral spin Seebeck effect, has been predicted in Ref. \cite{YuPRL2019} and attributed to the temperature difference between YIG and Py induced by the bias current flowing in Py.
This leads to the thermal excitation of magnons in the Py wire, which creates rotating dipolar stray fields with a chirality that depends on the Py magnetization direction. 
A magnon in Py can transfer its energy to a magnon in YIG as a function of its propagation direction through the coupling parameter $g_k$, resulting in a directional emission of magnons in YIG from the Py strip. This magnon generation mechanism does not occur through the exchange interaction between YIG and Py \cite{DasNL2018,DasPRB2020,SantosPRA2021}, as it is suppressed by the $\mathrm{TiO_x}$ layer. A large remote coupling parameter for a given $k$-vector results in an efficient magnon emission from the Py strip in a given direction, correlated with a low transmission under the Py strip of a magnon propagating in this same direction. This scenario is consistent with the results reported in Sections IV and V [summarized in Fig. \ref{fig:fig2}(e) and \ref{fig:fig2}(f)], indicating that our experimental results could be due to the chiral spin Seebeck effect.

\section{Conclusion}

We observed a non-reciprocity of the thermally-generated magnon transport in YIG, which depends on the direction of the associated temperature gradient, and on the magnetization direction of a Py strip placed between the heating and magnon detection points. The symmetries of this non-reciprocal signal indicate that it originates from a remote coupling between the magnon populations in YIG and in Py. This coupling is controlled by the temperature gradient and the Py magnetization direction, and originates from the chirality of the magnon dipolar stray fields. The effect of this coupling is also discussed for the transport of electrically-generated magnons. We stress the need to consider the specificities of near equilibrium transport processes for a proper description of the non-reciprocal magnon transport \cite{ HanNL2021,SchlitzPRL2021,GuckelhornPRL2023}.

In addition, we observed a directional emission of thermally-generated magnon in YIG from a Py strip. The directionality of the emission depends on the Py magnetization direction, and the dependence of the effect on the YIG magnetization direction indicates that it shares a common physical origin with the non-reciprocal magnon transport under the Py strip. This indicates that this observed directional magnon emission could correspond to the previously described \cite{YuPRL2019} chiral spin Seebeck effect.

\section*{Acknowledgment}
We acknowledge the technical support of J. Holstein, H.
de Vries, F. H. van der Velde, H. Adema, and A. Joshua. This work was supported by Zernike Institute for Advanced Materials (ZIAM), the Spinoza prize awarded to Professor B. J. van Wees by the Nederlandse Organisatie voorWetenschappelijk Onderzoek (NWO) in 2016, and has received funding from the European Research Council (ERC) under the European Union’s 2DMAGSPIN (Grant agreement No. 101053054).

%

\end{document}